\documentclass{cjpsuppl}
\usepackage{epsfig}

\usepackage{delarray,amsmath,amssymb,bbm}

\newcommand{\nn}{\nonumber}
\newcommand{\beq} {\begin{equation}}
\newcommand{\eeq} {\end{equation}}
\newcommand{\beqa} {\begin{eqnarray}}
\newcommand{\eeqa} {\end{eqnarray}}

\newcommand{\eg}{{\it e.g.}}

\newcommand{\eq}[1]{(\ref{#1})}

\newcommand{\prm}{\textrm{ .}}
\newcommand{\la}{\lambda}
\newcommand{\bs}[1]{\boldsymbol{#1}}
\newcommand{\vphi}{\varphi}
\newcommand{\im}{{\rm Im}}

\newcommand{\mM}{\mathcal{M}}

\newcommand{\pvec}{{\bs{p}}}

\newcommand{\qt}{q_\perp}
\newcommand{\pt}{p_\perp}

\newcommand{\lsim}{\lesssim}

\begin{document}
 \slacs{.6mm}
 \title{Single-spin asymmetry from pomeron-odderon interference}
 \authori{M. J\"arvinen}
 \addressi{Department of Physical Sciences and Helsinki Institute of
              Physics, POB 64, FIN-00014 University of Helsinki, Finland }
 \authorii{}    \addressii{}
 \authoriii{}   \addressiii{}
 \authoriv{}    \addressiv{}
 \authorv{}     \addressv{}
 \authorvi{}    \addressvi{}
 \headtitle{Single-spin asymmetry from pomeron-odderon interference}
 \headauthor{M. J\"arvinen}
 \lastevenhead{M. J\"arvinen: Single-spin asymmetry from pomeron-odderon interference}
 \refnum{}
 \daterec{} 
 \suppl{}  \year{2006} \setcounter{page}{1}
 \maketitle

 \begin{abstract}
The transverse single-spin asymmetry $A_N$ observed in
high energy proton-proton collisions $p^\uparrow p \to \pi X$ has been found to increase with the
momentum fraction $x_F$ of the pion up to the largest measured $x_F  
\sim 0.8$, where $A_N \simeq 40\%$. We consider the possibility that the asymmetry is due to 
a non-perturbatively generated spin-flip coupling in soft rescattering on the target proton.
We demonstrate using perturbation theory that a non-vanishing  
asymmetry can be generated through interference between exchanges of  
even and odd charge conjugation provided both helicity flip and non-flip  
couplings contribute. Pomeron and odderon exchange can thus explain the energy  
independence of the asymmetry and predicts that the asymmetry should persist in events with large rapidity gaps.
 \end{abstract}

\section{Introduction}

Large single spin asymmetries (SSA) have been observed in polarized hadron scattering $p^\uparrow p \to \pi(\pvec_\perp)X$ \cite{Adams:1991rw,Bravar:1996ki,Adams:2003fx} (as well as in $pp \to \Lambda^\uparrow X$ \cite{Lundberg:1989hw,Bravar:1995fw}). 
We consider the possibility that the large and energy independent asymmetries observed up to high values of of the energy fraction carried by the pion ($x_F \sim 0.8$) arise from pomeron-odderon interference. Previously, this mechanism has been suggested to cause charge and single spin asymmetries in diffractive $c\bar c$ \cite{Brodsky:1999mz} and $\pi^+\pi^-$ \cite{Ivanov:2001zc,Hagler:2002nh} photoproduction. A non-vanishing asymmetry arises if the odderon couples via a spin-flip coupling \cite{Kochelev:1996pv,Hoyer:2005ev}. Such a spin-flip coupling is absent in perturbative calculations, but might be generated non-perturbatively
due to QCD vacuum effects. This scenario can be experimentally verified by observing whether the asymmetry persists in events with large rapidity gaps.

The single spin asymmetry $A_N$ measures the dependence of the cross section on the spin direction of one particle polarized transverse to the reaction plane,
\beq \label{asymm}
 A_N \equiv \frac{\sum_{\{\sigma\}}\left[|\mM_{\uparrow,\{\sigma\}}|^2-|\mM_{\downarrow,\{\sigma\}}|^2 \right]}{\sum_{\{\sigma\}}\left[|\mM_{\uparrow,\{\sigma\}}|^2+|\mM_{\downarrow,\{\sigma\}}|^2  \right]} 
= \frac{2 \sum_{\{\sigma\}} \im\left[\mM_{\leftarrow,\{\sigma\}}^*\mM_{\rightarrow,\{\sigma\}}\right]}{\sum_{\{\sigma\}}\left[|\mM_{\rightarrow,\{\sigma\}}|^2+|\mM_{\leftarrow,\{\sigma\}}|^2  \right]}
\eeq 
where the $\mM_{\leftrightarrow,\{\sigma\}}$ are helicity amplitudes and $\{\sigma\}$ represents the helicities of all particles except the polarized one. Thus $A_N \neq 0$ requires two features which hard scattering amplitudes do not usually possess:
a helicity flip and a helicity-dependent phase. In our model, helicity flip is realized using a non-perturbative coupling
in soft (Pomeron, Odderon) scattering from the target. An analogous effect has previously been considered in \cite{Nachtmann:1983uz}. The helicity-dependent phase arises from interference between pomeron and odderon exchange. 

\section{A perturbative model}
 
We study first the SSA mechanism in $p^\uparrow p \to \pi(\pvec_\perp) X$ using an abelian gluon exchange model, namely $e^\uparrow \mu \to e(\pvec_\perp)\gamma \mu$ where only the momentum of the electron is measured in the final state (Fig.~1). In QCD, the photon and the electron would be replaced by a gluon and a quark which fragments into a pion. The electron-photon vertex is the hard part which generates the large $\pt$ of the pion. Since the data indicate that $A_N$ is independent of the center-of-mass energy, we take the limit $E_{CM} \to \infty$ at a fixed (and large) $\pt$ of the photon emitted from the electron. At leading order in $E_{CM}$ only soft (Coulomb) photon exchange between the projectile (electron) and target (muon) systems contributes. The electron will be referred to as ``quark'' and the photons as ``gluons'' in the following.

\begin{figure}[thb]
 \begin{center}
 \includegraphics[width=0.7\textwidth]{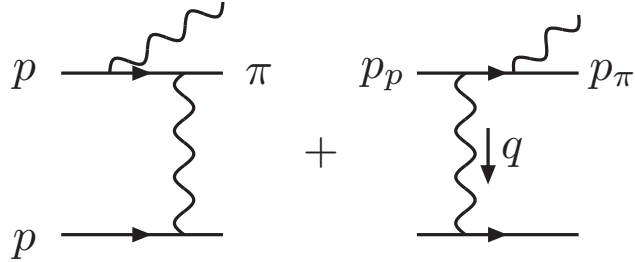}
 \end{center}
 \caption{The Born diagrams in our model.}
\end{figure}

It is straightforward to verify that in the model described above soft Coulomb rescattering does not contribute to $A_N$: adding multiple Coulomb exchanges to the diagrams of Fig.~1 only gives a helicity 
{\em independent} Coulomb phase that cancels in \eq{asymm}. However, a spin-flip coupling at the rescattering vertex can lead to helicity dependent phases and nonzero asymmetries \cite{Kochelev:1996pv,Hoyer:2005ev}.
Such a coupling is generated perturbatively by a gluon loop correction to the pointlike vertex. However, since the dominant contribution comes from low transverse momenta the formation time $\tau$ of the gluon-quark pair in such a loop is long compared to the hard gluon emission required to generate the transverse momentum of the pion (in the target rest frame $\tau \sim 1/\pt\cdot s/M\pt$ where $M$ is the target mass). Hence coherence is lost and
the perturbative contribution expected to be supressed.
Nevertheless, for soft gluon exchange the coupling may be generated by the non-perturbative, chirality breaking sector of QCD, to which above timing argument does not apply \cite{Nachtmann:1983uz}.

We study the Pauli coupling which is defined by replacing
\beq \label{paulivert}
 -ie\gamma ^\mu \to -ie\gamma^\mu + a(q^2)\sigma^{\mu\nu}q_\nu \prm
\eeq 
The effective coupling $a(q^2)$ in \eq{paulivert} should vanish at large virtualities $-q^2$ of the gluon when the vertex becomes perturbative. The precise form of $a(q^2)$ is not important. We use
\beq
 a(q^2) = a_0 \exp(-A\qt^2) \prm
\eeq

Our model with spin-flip rescattering is as described above ($e^\uparrow \mu \to e(\pt)\gamma \mu$) but with 
a possibility of having a Pauli coupling at the soft gluon vertex. We fix the frame such that 
\beqa
 p_p&=&(p^+,m^2/p^+,0,0)\nn\\
 p_\pi&=&(x_F\, p^+,p_\pi^-,\pt\cos\vphi,\pt\sin\vphi)\nn\\
 q&\simeq&(0,0,\qt\cos\phi,\qt\sin\phi)
\eeqa
where $p_p$, $p_\pi$ and $q$ are the incoming quark, outgoing quark (pion) and exchanged soft gluon momenta, respectively (see Fig.~1). 
We drop the small terms which result from spin flips at vertices other than the Pauli vertex by taking the quark mass $m$ to zero. Some of the diagrams needed in the calculation are shown in Fig.~2. A lengthy calculation gives the asymmetry, which reads for soft rescattering $\qt^2 \sim 1/A \ll \pt$ and $a_0 \pt \ll 1$
\beq \label{perres}
 A_N \propto -\frac{ e a_0\ x_F\ \pt }{1+x_F^2}\cos \vphi \prm
\eeq

\begin{figure}[thb]
 \begin{center}
 \includegraphics[width=0.7\textwidth]{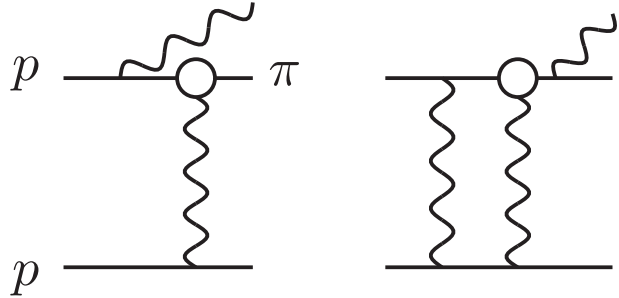}
 \end{center}
 \caption{Examples of Born and loop diagrams needed when evaluating $A_N$ to first order in the spin-flip coupling ($a_0$, the blob). In total, there are four Born diagrams and nine loop amplitudes with discontinuities (absorbtive parts).}
\end{figure}

The flip amplitudes with Pauli coupling are $\propto 1/\pt$, while the non-flip amplitudes behave as $1/\pt^2$ for soft rescattering $\qt \ll \pt$. Thus the asymmetry \eq{perres} is $\propto \pt$ in the region where the non-flip amplitudes dominate ($a_o \pt \ll 1$).
The $x_F$ and $\pt$ dependencies of \eq{perres} arise from the hard gluon emission vertex. The soft Coulomb exchange only affects the proportionality constant: it is given by a ratio of soft integrals and its magnitude depends of the size of the phase between the flip and non-flip amplitudes. The phase turns out to be small since the contributions from the (imaginary) double Coulomb exchange diagrams are smaller than the (real) Born contributions. Hence a rather small asymmetry ($A_N\lsim 5\%$) is generated.

\section{Asymmetry from pomeron-odderon interference with a helicity flip}

We shall now use the ideas of the previous section to build a model with Regge amplitudes for the pomeron
and odderon exchange. In this model the dynamical phase is not generated by rescattering but through an interference between the pomeron and odderon exchange amplitudes \cite{Brodsky:1999mz}. If the large asymmetries observed in $p^\uparrow p \to \pi X$ are to be explained using pomeron-odderon interference, the phase needs to be $\sim 90^\circ$. The pomeron spin-flip amplitudes are observed to be small (see, \eg, \cite{Okada:2006dd}).
The odderon may on the other hand have a sizeable spin flip coupling. This would provide the required large phase difference between the spin and non-flip parts of the total amplitude due to the different signatures of the pomeron and odderon.

We use for the pomeron (odderon) exchange amplitudes the Regge form
\beq \label{reggeamp}
 (i) \exp\left(\frac{i \pi \alpha_{P/O}}{2}\right) \exp\left(bt\right)\left(\frac{s}{s_0}\right)^{\alpha_{P/O}}
\eeq
where the factor $i$ appears only in the odderon amplitude as a consequence of the negative signature of the odderon. For our purposes the $t$ dependencies of $\alpha_P$ and $\alpha_O$ can be neglected. We also approximate $\exp(i \pi \alpha_{P/O}/2) \simeq i$ as we will have $\alpha_P \simeq \alpha_O \simeq 1$.

\begin{figure}[thb]
 \begin{center}
 \includegraphics[width=0.7\textwidth]{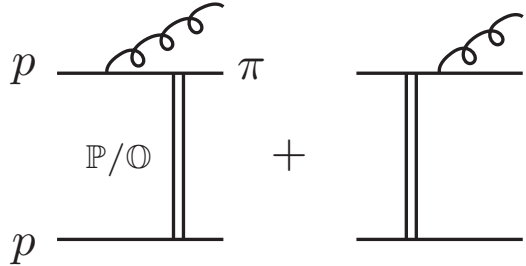}
 \end{center}
 \caption{Diagrams of the minimal model with pomeron-odderon interference. The odderon coupling is assumed to flip spin.}
\end{figure}

We consider a minimal model with the two diagrams shown in Fig.~3. For the hard vertex we use the QED Feynman rules, whereas 
the amplitude of the soft exchange is given by the Regge form \eq{reggeamp}. The amplitudes with pomeron exchange become (for $\qt \ll \pt$)
\beq
 \mM^P_{++\la} \simeq i\la x_F^{(1-\la)/2}e\sqrt{2 x_F}(1-x_F)  e^{bt}\frac{s}{s_0}\frac{\qt e^{i\phi\la}}{\pt^2 e^{2i\varphi\la}}
\eeq
where we fixed $\alpha_P=1$.

The odderon contribution is assumed to flip spin:
\beq
 \mM^O_{+-\la} \simeq -\la e a_0 \sqrt{\frac{2}{x_F}} \qt e^{i\phi} e^{bt} \left(\frac{s}{s_0}\right)^{\alpha_O}
\frac{x_F^{(3+\la)/2} - x_F^{\alpha_O+(1-\la)/2}}{\pt e^{i\varphi\la}}
\eeq
where $a_0$ is the strength of the spin-flip coupling and the phase $e^{i\phi}$ is required by the rotational properties of the amplitude (angular momentum conservation).

\begin{figure}[thb]
 \begin{center}
 \includegraphics[width=0.75\textwidth]{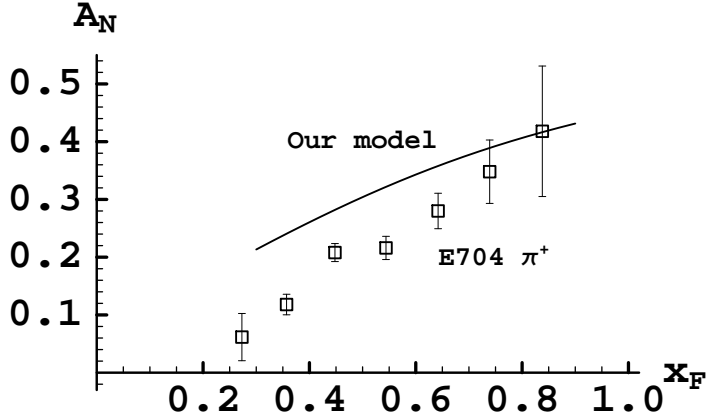}
 \end{center}
 \caption{$A_N$ resulting from pomeron-odderon interference. The solid line is our numerical result, and the boxes show the E704 data for $A_N$ in $p^\uparrow p \to \pi^+ X$ \cite{Adams:1991rw}.}
\end{figure}

Using $\alpha_O = \alpha_P =1$ and $\qt \ll \pt $ the resulting asymmetry reads
\beq \label{pores}
 A_N \simeq -\frac{2 a_0 \pt x_F}{1+x_F^2+2 a_0^2\pt^2}\cos\vphi \prm
\eeq
It increases with $x_F$ and with $\pt$ up to $\pt \sim 1/a_0$ in accordance with the data. Similarly to \eq{perres} the $x_F$ and $\pt$ dependencies of \eq{pores} arise from the hard vertex. In general, the asymmetry can be found by numerical integration. Fig.~4 shows $A_N$ for $\alpha_P=1$, $\alpha_O=0.9$, $a_0(s/s_o)^{\alpha_O-\alpha_P}=-0.4 $ GeV$^{-1}$, $b=7$ GeV$^{-2}$ and assuming a correlation $\pt = (0.2+ 0.8 x_F)$ 
GeV\footnote{$x_F$ and $\pt$ are similarly correlated in the experimental data \cite{Bravar:1996ki}.}.

 \section{Conclusion}

We suggested a novel mechanism to explain the large single spin asymmetry observed in $p^\uparrow p \to \pi X$. 
A pertubative calculation motivated us to consider the possibility that the asymmetry is due to pomeron-odderon interference.  
A spin-flip odderon coupling provides the helicity-dependent phase which is required for a sizeable single spin asymmetry. Note that if the 
asymmetry is created via pomeron/odderon exchange it should persist in events with large rapidity gaps. It should be possible to experimentally
test this prediction.

 \bigskip

 {\small I would like to thank the organisers of Spin-Praha-2006 for invitation to the conference and my supervisor Paul Hoyer for his advice. Also thanks to the participants of Spin-Praha-2006 for a number of interesting comments and to Oleg Selyugin for a useful discussion. This research was supported in part by GRASPANP, the Finnish Graduate School in Particle and Nuclear Physics, and by the Academy of Finland through grant 102046.}


 \end{document}